\documentclass[manuscript,screen,review]{acmart}

\usepackage[official]{eurosym}
\AtBeginDocument{%
  \providecommand\BibTeX{{%
    \normalfont B\kern-0.5em{\scshape i\kern-0.25em b}\kern-0.8em\TeX}}}
\setcopyright{acmcopyright}
\copyrightyear{2021}
\acmYear{2021}
\acmDOI{10.1145/1122445.1122456}
\acmConference[]{}{}{}
\acmBooktitle{}
\acmPrice{}
\acmISBN{}

\usepackage{multirow}

\begin{document}

%%
%% The "title" command has an optional parameter,
%% allowing the author to define a "short title" to be used in page headers.
\title{LookAtChat: Visualizing Eye Contacts for Remote Small-Group Conversations}

%%
%% The "author" command and its associated commands are used to define
%% the authors and their affiliations.
%% Of note is the shared affiliation of the first two authors, and the
%% "authornote" and "authornotemark" commands
%% used to denote shared contribution to the research.
\author{Zhenyi He}
\affiliation{%
  \institution{New York University}
  \streetaddress{60 5th Ave}
  \city{New York}
  \country{United States}}
\email{zhenyi.he@nyu.edu}
% \orcid{1234-5678-9012}

\author{Ruofei Du}
\email{me@duruofei.com}
\affiliation{%
  \institution{Google}
  \country{United States}
}

\author{Ken Perlin}
\affiliation{%
  \institution{New York University}
  \streetaddress{60 5th Ave}
  \city{New York}
  \country{United States}}
\email{perlin@cs.nyu.edu}

\newcommand{\etal}{{\em et al. }}

\begin{abstract}
% 150 words limit

% 150

% LookAtChat
Video conferences play a vital role in our daily lives. However, many nonverbal cues are missing, including gaze and spatial information. We introduce LookAtChat, a web-based video conferencing system, which empowers remote users to identify eye contact and spatial relationships in small-group conversations. Leveraging real-time eye-tracking technology available with ordinary webcams, LookAtChat tracks each user's gaze direction, identifies who is looking at whom, and provides corresponding spatial cues. Informed by formative interviews with 5 participants who regularly use videoconferencing software, we explored the design space of eye contact visualization in both 2D and 3D layouts. We further conducted an exploratory user study (N=20) to evaluate LookAtChat in three conditions: baseline layout, 2D directional layout, and 3D perspective layout. Our findings demonstrate how LookAtChat engages participants in small-group conversations, how gaze and spatial information improve conversation quality, and the potential benefits and challenges to incorporating eye contact visualization into existing videoconferencing systems.

\end{abstract}

%%
%% The code below is generated by the tool at http://dl.acm.org/ccs.cfm.
%% Please copy and paste the code instead of the example below.
%%


%%
%% Keywords. The author(s) should pick words that accurately describe
%% the work being presented. Separate the keywords with commas.
\begin{teaserfigure}
  \includegraphics[width=\textwidth]{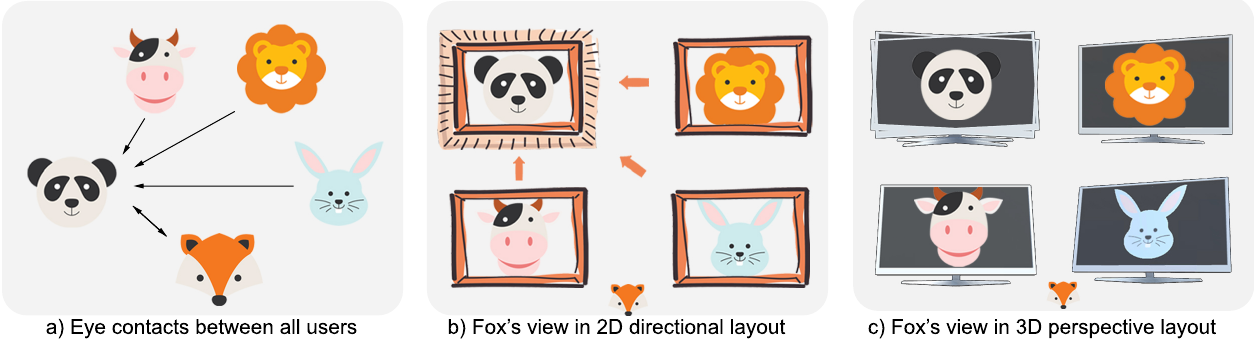}
%   a) Five in a circle, b) 2D, glow and arrow, c) 3D, shake and turning
  \caption{We present \textit{LookAtChat}, an online group chat system that takes advantage of eye-tracking technology available with ordinary webcams to visualize eye contacts in group chats. LookAtChat implements and evaluates both 2D directional and 3D perspective layout to inform users with spatial cues. To convey eye contacts in a),  b) directional layout renders arrows between video streams, while out-glowing the video window of users who are looking at you. c) Perspective layout transforms video streams to gaze targets, while slightly shaking the video window of users who are looking at you. As a proof-of-concept, LookAtChat explores the potential of using gaze information during online video chat.}
%   a)b)c)
%   \Description{FOR ACCESSIBILITY.}
  \label{fig:teaser}
\end{teaserfigure}

%%
%% This command processes the author and affiliation and title
%% information and builds the first part of the formatted document.
\maketitle

\section{Introduction}

Video conferencing is becoming the dominant medium for remote discussion and collaboration during the global COVID-19 pandemic. However, \textit{eye contact information for more than two people} in mainstream video chat tools such as Skype, Zoom, and Google Meet is insufficiently conveyed through this medium. Hence, it is almost impossible to use eye gaze as a nonverbal cue to infer where attention is directed in conventional video conferences.

Prior art in remote small-group collaboration has leveraged multi-view cameras, customized displays, or mixed-reality settings to solve this problem. For example, GAZE-2 \cite{vertegaal1999gaze, vertegaal2003gaze} employs an array of cameras with an eye tracker and selectively transmits the preferred video stream to remote users. MMSpace \cite{otsuka2016mmspace} introduces novel physical kinetic displays to support eye contacts between every pair of participants; Holoportation \cite{orts2016holoportation} leverages full-body reconstruction and headset-removal technologies to achieve immersive small-group telepresence with Microsoft HoloLens. However, it is still unclear how to embed eye contact in a conventional videoconference with an ordinary laptop; or to determine the potential benefits and drawbacks of visualizing eye contacts in remote small-group conversation.

In this paper, we present and evaluate LookAtChat, a video conferencing system which visualizes eye contacts for remote small-group conversation. LookAtChat consists of three components: a WebRTC server to support videoconferencing and logging, an eye-tracking module powered by \verb+WebGazer.js+ \cite{papoutsaki2016webgazer} to recognize gaze positions, and a visualization module implemented with the \verb+three.js+\footnote{three.js: JavaScript 3D library, \url{http://www.threejs.org}.}. 

As initial work, our research questions are exploratory: How do people perceive eye contact in conventional video conferences? Can visualization of eye contact improve remote communication efficiency? In what context will users prefer to see eye contact? What forms of eye-contact visualization may be preferred by users? 

To create LookAtChat, we conducted formative interviews with five people who use videoconferencing with colleagues on a daily basis. Our research is inspired and informed by prior small-group communication systems that demonstrate the potential of visualizing eye contacts and spatial information. To improve the generalizability and replicability of the system, we only require each user to use a laptop with a webcam. We further extend the design space of visualizing gaze and spatial information to a total of 11 layouts. % Our research is inspired and informed by prior small-group communication systems that demonstrate the potential of visualizing eye contacts and spatial information. We extend the research in three ways.

To evaluate LookAtChat, we conducted four three-session user studies with 20 remote participants (ages 24-39, 6 female and 14 male). In our analyses of video recordings, post-activity questionnaires, and post-hoc interviews, we found that LookAtChat can effectively engage participants in small-group conversations by visualizing eye contact and providing spatial relationships. Gaze and spatial information can improve the conversation experience, bring greater social presence and richness, and provide better user engagement.
Our main contributions in this paper are: 

\begin{enumerate}
% design space
    \item Conception, development, and deployment of LookAtChat, a video conferencing system that can visualize eye contact for remote small-group conversations.
    \item Enumerating design implications through formative interviews and extending the design space of visualizing gaze and spatial information in video conferences.
    \item Reporting evaluation results and reflections about the opportunistic use of eye contact visualization in video conferencing systems - benefits, limitations, and potential impacts to future remote collaboration systems.
    \item Open-sourcing\footnote{The Github \href{maskForAnonymous}{repository} containing the source code will be available in the camera-ready revision. A live demo is available at \url{https://eye.3dvar.com}}. We plan to make our software available to facilitate future development in video conferencing systems with visualization of nonverbal cues.
\end{enumerate}

% Names:
%
% LookAtChat
% GazeChat
% EyeChat
% GazeChat
% Vis
% Video
% 

% Backburners:
% commercial platforms such as VRChat and Second Life use virtual avatars to direct attention,
% it is almost impossible to infer eye contacts in the mainstream video conferencing software for a call with more than two people, 
% as users' gaze directions are mapping into different parts of the display 

% ~\cite{regenbrecht2015mutual}
% cues are highly important for effective communication.
% Mutual gaze support plays a central role in those high coordination need scenarios but generally lacks adequate technical support from videoconferencing systems

\section{Related Work}
% TODO:summary at the beginning of each paragraph

To understand how gaze information is integrated into video conferences and to justify our design decisions, we review prior art on multi-user experience in distributed collaboration and gaze tracking technologies for videoconferencing. Many researchers have contributed to investigating future workspaces such as improving individual productivity like HoloDoc~\cite{li2019holodoc}, reconstructing multi-user experiences like ``the office of the future''~\cite{raskar1998office}, and cross-device interaction~\cite{voelker2020gazeconduits}. Furthermore, remote conferencing shows its potential for geologically dispersed users and is efficient for group discussion~\cite{neustaedter2015sharing}. In scenarios requiring certain levels of trust and judgement with non-verbal communication, non-verbal cues are highly important for effective communication~\cite{regenbrecht2015mutual}. Gaze support and feeling of face-to-face~\cite{olson1995mix} play a central role in those scenarios. With the increasing development of gaze tracking devices and technology, gaze-assisted interaction are becoming popular in the fields of text entry~\cite{ward2000dasher}, video captions~\cite{kurzhals2020view}, and video conferences. 

\subsection{Multi-user Collaboration in Distributed Environments}

Distributed multi-user collaboration has been widely researched from the perspective of locomotion, shared proxies, and life-size reconstruction as well as different purposes including communication, presentation, and object manipulation.
{Your Place and Mine}~\cite{sra2018your} creates experiences that allow everyone to real walk in collaborative VR.
{Three's Company}~\cite{tang2010three} presents a three-way distributed collaboration system that places remote users either on the same side or around a round table. In addition, Three's Company provides non-verbal cues like body gestures through a shared tabletop interface. Remote users' arm shadows are displayed locally on a tabletop device, which is beneficial for collaborative tasks with shared objects.
Tan~\etal~\cite{gazeAwareness} focus on presentation in large-venue scenarios, creating a live video view that seamlessly combines the presenter and the presented material, capturing all graphical, verbal, and nonverbal channels of communication.
Tele-Board~\cite{gumienny2011tele} enables regionally separated team members to simultaneously manipulate artifacts while seeing each other’s gestures and facial expressions.
The concept of {Blended Interaction Spaces}~\cite{o2011blended} is proposed to providing the illusion of a single unified space by creating appropriate shared spatial geometries.
{TwinSpace}~\cite{reilly2010twinspace} is a generic framework discussing brainstorming and presentation in cross-reality that combines interactive workspaces and collaborative virtual worlds with large wall screens and projected tabletops.
{Physical Telepresence workspaces}~\cite{leithinger2014physical} is a shaped display providing shape transmission that can manipulate remote physical objects.
Cameras are widely used for above alternatives to capture users, besides, 360 video has recently been researched.
{SharedSphere}~\cite{lee2018user} is a wearable MR remote collaboration system that enriches a live captured immersive panorama based collaboration through MR visualisation of non-verbal communication cues.

Immersive collaborative virtual environment (ICVE) and Augmented Reality (AR) can be used to develop new forms of teleconferencing, which often leverages multiple cameras setup and 3D reconstruction algorithms.
% Providing eye-gaze immersive collaborative virtual environment (ICVE) systems is under discussion for decades.
{EyeCVE}~\cite{steptoe2008eye} uses mobile eye-trackers to drive the gaze of each participant's virtual avatar, thus supporting remote mutual eye-contact and awareness of others' gaze in a perceptually coherent shared virtual workspace.
Jones~\etal~\cite{jones2009achieving} design a one-to-many 3D teleconferencing system able to reproduce the effects of gaze, attention, and eye contact. A camera with projected structure-light is set up for reconstructing the remote user.
Billinghurst and Kato~\cite{billinghurst2000out} developed a system that allows virtual avatars and live video of remote collaborators to be superimposed over any real location. Remote participants were mapped to different fiducial markers. The corresponding video images were attached to the marker surface when markers are visible.
{Room2Room}~\cite{pejsa2016room2room} is a telepresence system that leverages projected AR to enable life-size, face-to-face, co-present interaction between two remote participants by performing 3D capture of the local user with RGBD cameras.
{Holoportation}~\cite{orts2016holoportation} demonstrates real-time 3D reconstructions of an entire space, including people, furniture and objects, using a set of depth cameras. Gestures are preserved via full-body reconstruction and headset removal algorithms are designed to convey eye contact. However, ``uncanny valley'' remains a challenging problem in this domain.

\subsection{Eye Contacts and Gaze Correction Technology in Video-mediated Conversation}

Various hardware setups have been explored for gaze correction including hole in screen, long distance, and half-silver mirror. The hole in screen concept is about drilling a hole in the screen and placing a camera. Long distance uses a screen at a far distance while placing the camera as close as possible~\cite{tam2007perception}. Half-silver mirror allows a user to see through a half-transparent mirror while being observed by a well-positioned camera at the same time. This idea was adapted in {ClearBoard}~\cite{harrison1995transparent, ishii1992clearboard} and Li~\etal's transparent display~\cite{li2014interactive}. Despite their advantages in terms of system complexity and costs, such solutions are rarely used outside of laboratory due to the availability of hardware.
In the meantime, quite a few 2D video-based (or image-based) approaches are proposed for eye contact including eye correction with a single camera~\cite{andersson1996video,andersson1997method} and multiple cameras~\cite{criminisi2003gaze} while applying image-based approaches like texture remapping and image warp~\cite{gemmell2000gaze}. However, the technology is not sufficiently accurate to avoid visual artifacts and the uncanny valley. 3D video-based solutions including 3D reconstruction is another trend for maintaining eye contact while the head is reconstructed. RGB camera~\cite{xu1999true}, depth camera~\cite{zhu2011eye}, Kinect~\cite{kuster2012gaze}, or motion capture system~\cite{maimone2013general} are used for 3D reconstruction. 

Eng~\etal~\cite{eng2013gaze} propose a gaze correction solution for a 3D teleconferencing system with a single color/depth camera. A virtual view is generated in the virtual camera location with hole filling algorithms.
Compared to single camera setup, multiple cameras are popularly used for providing gaze~\cite{ashdown2005combining} in videoconferencing.
True-view~\cite{xu1999true} was implemented with two cameras (one on the left and the other on the right). The synthesised virtual camera view image at the middle viewpoint is generated to provide correct views of each other and the illusion of close proximity.
GAZE-2~\cite{vertegaal1999gaze, vertegaal2003gaze} utilizes an eye tracker with three cameras. The eye tracker is used for selecting a proper camera closest to where the user is looking. GAZE-2 prototypes an attentive virtual meeting room to experiment with camera selection. In each meeting room, each user’s video image is automatically rotated in 3D toward the participant he is looking at. All the video images are placed horizontally so the video image turns left or right when the corresponding camera is chosen.
Likewise, {MultiView}~\cite{nguyen2005multiview, nguyen2007multiview} is a video conferencing system that supports collaboration between remote groups of people with three cameras. Additionally, {MultiView} allows multiple users to be co-located in one site by generating a personal view for each user even though they look upon the same projection surface, which they achieve by using a retro-reflective material.
{Photoportals}~\cite{beck2013immersive, kunert2014photoportals} groups local users and remote users together through a large display. All users are tracked and roughly reconstructed through multiple cameras and then rendered within a virtual environment.
{MMSpace}~\cite{otsuka2012reconstructing, otsuka2013mm+, otsuka2016mmspace, otsuka2017behavioral} provided realistic social telepresence in symmetric small group-to-group conversations through ``kinetic display avatars''. Kinetic display avatars can change pose and position by automatically mirroring the remote user’s head motions. One camera is associated with one transparent display. Both camera and display can be turned to provide corresponding video input image and output angle.
Sirkin~\etal~\cite{sirkin2011motion} developed a kinetic video conferencing proxy with a swiveling display screen to indicate which direction in which the satellite participant was looking for maintaining gaze and gestures to mediate interaction.
Instead of rendering a video image on a rectangular display, a cylinder display is proposed in {TeleHuman}~\cite{kim2012telehuman} with 6 Kinects and a 3D projector.

% \subsection{???}
% GazeConduits~\cite{voelker2020gazeconduits} is a calibration-free ad-hoc mobile interaction concept that enables users to collaboratively interact with tablets and content in a cross-device setting using gaze and touch input. Gaze tracking is implemented on the top of facial features provided by modern mobile devices.

LookAtChat is designed to be used with \textit{a minimum requirement of a laptop/PC and a single webcam}. While multi-view cameras and external hardware may yield better eye tracking and 3D rendering solutions, such systems typically require very high computational power and exclusive hardware setups. Since it is possible for users with low-cost video conferencing setup to learn to interpret gaze direction to a very high degree of accuracy~\cite{grayson2003you}, we decided not to apply extensive image-based manipulation on video streams but rather to focus on the design of a widely accessible online system to empower video conferencing users with real-time visualization of eye contacts.

\section{Formative Interviews}
% user research https://docs.google.com/document/d/1AyvcTYGokSX8d_wBc7f2Pq9ZvNuLXpqilee8glb2o2M/edit?usp=sharing
% interview videos https://drive.google.com/drive/folders/1uaT_q2s-u2yIFBi_7LEMa4QkS759M8Fc?usp=sharing

% find example formative interview
% describe the interview we conducted with several people to understand how people think about eye contact in video conferencing
% summarize the challenge, concerns, and potential feature we want to add for our study
% this section is optional or could be part of the design

% create a doc including interview notes

To inform the design of LookAtChat and understand whether and how gaze information affects video conferencing, we conducted five formative interviews with videoconferencing users (2 female and 3 male, labeled as I1 to I5) to learn the advantages and disadvantages of current videoconferencing software compared to real-life meetings as well as people's expectation of videoconferencing. We asked participants about their recent video conferencing experience under different scenarios. Our takeaways are summarized below.
%TODO: what different scenarios.
% small group, large scale, with presentation, with whiteboarding

\subsection{Feedback on the Existing Videoconferencing Systems}

\textbf{Good for multi-tasking and information sharing.}

Software such as \textit{Zoom} and \textit{Skype} allows participants to work on multiple tasks at the same time while video conferencing, such as walking on a treadmill while listening to a talk. Users benefit from sharing screen or notes through videoconferencing software, as it allows any participant to instantly share their own document or presentation. Although participants in offline meetings can share information through whiteboarding or printed documents, video conferencing software allows a large number of people to concentrate on the same document and work on different sections of it.

\textbf{Bad for white-boarding and body gestures.}

For group discussions in which all participants may need to contribute their thoughts, a physical whiteboard is very popular. And yet, shared free sketch software is not well integrated into video conferencing software or available as stand-alone software for now, though quite a few researches have focused on that in immersive environments. Similarly, body gestures are partially missing due to the small view area of cameras and missing/different spatial information of participants.

\textbf{Bad for finding the speaking up timing.}

P1 and P4 thought it was more difficult to know when to speak in online meetings because not all the participants' gaze and body information are perceived well through the camera. It is not clear who is talking to whom, whether the speaker is waiting for an answer from a specific person, or if a speaker is pausing or is ending the conversation during group discussion.

\textbf{Bad for controlling meeting length.}% TODO

The length of physical meetings are usually well controlled since the meeting rooms are usually booked throughout the day and participants are aware of those who are gazing through the window, waiting to use the room next. However, participants in virtual conferences often cannot find the best time to exit for the next meeting. I2(M) commented that ``in virtual video conferences, very few people strictly follow the proposed length of the meeting and oftentimes delay the next meetings. People just keep talking when the meeting goes beyond the scheduled time''.

\subsection{Expectations of Future Videoconferencing Systems}

\textbf{Improve the control of conversation.} It is difficult to use words such as ``you'' in video conferencing contexts because participants barely know who the speaker is talking to, while ``you'' is natural in co-located conversations. I1(M) felt ``less involved'' because of the lack of this information. There also exists more simultaneous speech in video conferences. People start talking together and stop together to wait, which causes participants to lose track of the conversation.

\textbf{Provide spatial information.}
I3(F) wished to select a seat the way they would normally enter a meeting room: ``Everyone has their own perspectives and maintain spatial relationship with each other''.

\subsection{Thoughts of Visualizing Gaze Information in Videoconferencing System}

\textbf{Good for natural discussion.} Interviewee (I2(M) and I5(F)) think it is helpful especially if the discussion requires feedback, attention, and interaction. Also, it is helpful for branching ideas. It is easy to suggest what topic one participant is following by looking at the proposer directly in offline meetings, but not easy to show to the group information in videoconferencing software.

\textbf{Different for small group and large scale.} For presentations or lectures, presenters or teachers may benefit from participants' gaze information that helps them adjust content in real-time. P1 elaborated: ``teachers know the topic is difficult or get distracted when quite a few students' gaze focuses are shifted.''

\textbf{Concerns for privacy.} Some interviewees (I3(F) and I5(F)) mentioned that they feel pressured when being looked at or looking at others. Displaying anonymous gaze information or aggregated data and reporting the result afterwards may be helpful.

\section{LookAtChat}
Informed by formative interviews and inspired by prior systems, we formulate our design rationale, explore the design space, elaborate on two specific layouts for natural integration with conventional videoconferencing, and discuss potential use cases.

\subsection{Design Rationale}
% Inspired by https://drive.google.com/file/d/1M5PAuI1wMYVr_JNeDs-TDNNZbQB-eo6A/view?usp=sharing
% Pronto: Rapid Augmented Reality Video Prototyping Using Sketches and Enaction
% define our vc
We constrain our design scope to remote small-group conversations in which all participants are physically dispersed. This setting is motivated by the circumstances of COVID-19, where everyone is working remotely.
Users mostly participate in these conversations on a laptop with a built-in frontal camera, or a workstation with a USB webcam.
Scenarios with two or more people co-located in front of the camera for video conferencing are out of our design scope.
Depth camera\cite{pejsa2016room2room}, multiple cameras\cite{orts2016holoportation}, motion capture systems\cite{maimone2013general}, professional eye-tracking systems\cite{vertegaal2003gaze}, or head-mounted displays\cite{orts2016holoportation} are not considered as alternatives in our design due to their constraints of cost and availability.
Although the above devices allow richer social engagements and more accurate gaze detection over a single webcam, we desire to make our platform accessible to most users.
Taking these factors into account, we constructed a web application to prototype LookAtChat so that any device with a camera can access our website via a modern Internet browser such as Google Chrome. 

\subsection{Design Space}

\begin{figure}[t]
    \centering
    \includegraphics[width=\textwidth]{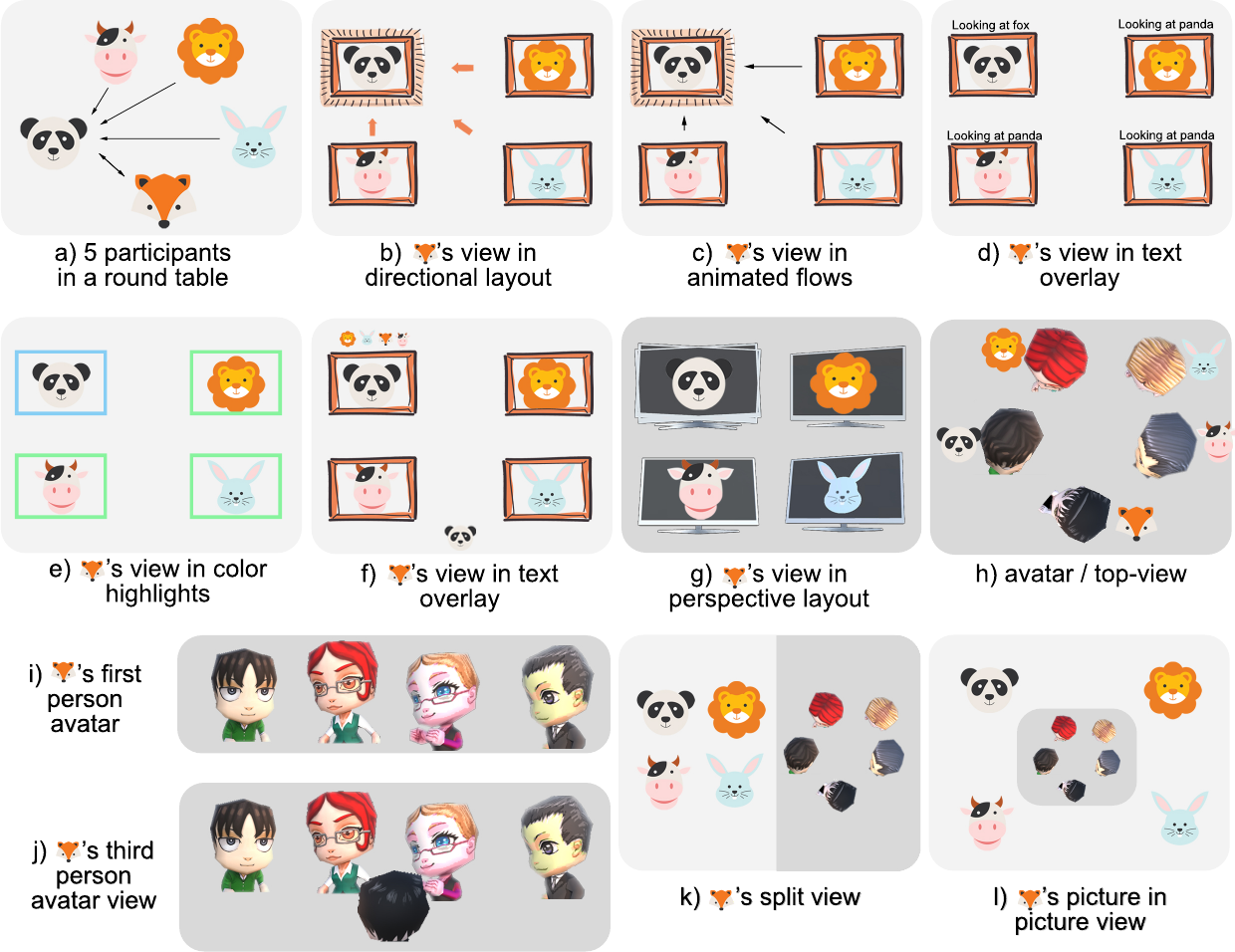}
    \hfill
     \caption{Design space of LookAtChat. To convey eye contacts in a), 2D flat layouts (b -- f), 3D immersed layouts (g - j), and hybrid layouts (k -- l) illustrate the mutual gaze between ``panda'' and ``fox'' and how other participants gaze at ``panda''.}\label{fig:designspace}
\end{figure}

Elicited from formative interviews, we prototype LookAtChat to explore the design space visualizing gaze in videoconferencing. Considering popular video conferencing software is using 2D flat layouts for video image placement, we explore the design space on top of the traditional 2D flat layouts in order to expand the potential of 3D as well as hybrid layout alternatives. Hybrid dimension alternatives are proposed to combine 2D and 3D representations for taking advantage of both categories. The short description of our designs are listed in \autoref{tab:designspace} with corresponding illustrations in \autoref{fig:designspace}. Each sub-figure in \autoref{fig:designspace} illustrates a five-user scenario: user ``panda'' is speaking and looking at user ``fox'' while all the rest participants are listening to ``panda'' and looking at ``panda''.

% uncanny
\begin{table}[htb]
  \caption{Proposed visualization of eye contact for remote video conferences}
  \label{tab:designspace}
  \begin{tabular}{p{6em} p{8.5em} p{29em}}
    \toprule
    Category & Name & Description\\
    \midrule
    2D flat layout & directional layout    & Depict arrows between video streams to indicate sources and targets of eye contacts, while out-glowing the video window of users who are looking at the current user.\\
                                    & animated flows        & Render dynamic flows instead of arrows to convey eye contacts; the sizes of the flows are proportional to the duration of the gaze actions.\\
                                    & text overlay          & Overlay the text of ``[who] is looking at [whom]'' directly on captions of the video window. \\
                                    & color highlights      & Change the color of the video border to indicate the eye contacts while each distinct color is assigned to each participant. \\
                                    & icon overlay          & Append users' profile pictures to the caption area of the video window to convey eye contacts.\\
    \midrule
    3D immersed layout
        & perspective layout        &  Apply perspective transformation of the video window to imply eye contact between other participants and gently shake the video of users who are looking at the current user. \\
        & avatar / top-view         & Render a top-view of 3D avatars of all users alongside with their video streams and change their rotation according to gaze actions. \\ % 3DHumanModel + different texture
        & avatar / first-person     & Warp live video streams to the 3D avatars of all users positioned along a curve; rotate the avatars to reflect their gaze actions; present a first-person perspective for the current user\\
        & avatar / third-person     & Based on ``avatar / first-person'', present a third-person perspective with isometric projection\cite{cai2007isometric}.\\ % https://en.wikipedia.org/wiki/Isometric_video_game_graphics
    \midrule
    hybrid layout   & split-view            & Present a 2D flat layout and a 3D immersed layout side by side. Hence the 3D avatar layout doesn't need to wrap video streams to the avatars to avoid ``uncanny valley'' effects.\\% and allow users to switch naturally in between.
                    & picture-in-picture    & Depict a 2D flat layout with video windows in full while a 3D avatar layout is rendered as an overview thumbnail at the screen center to convey eye contacts.\\
  \bottomrule
\end{tabular}
\end{table}

% emphasize and neglect
\subsubsection{2D flat layout}

We first explore how to impose eye contact on 2D flat layout. The visualization of eye contact on 2D flat layouts could be illustrated in a direct or indirect manner. Direct design delivers straightforward signals with less cognitive load, while indirect design may result in less interference with the video streams. \autoref{fig:designspace}(b) and (c) are illustrated how \textit{directional layout} and \textit{animated flows} convey eye contact. From perspective of user `fox', `panda' is looking at itself so the video frame is highlighted with outer glows. In the meantime, all other participants are looking at `panda' so a static directional arrow is shown in \autoref{fig:designspace}(b) and dynamic flow from observer to observee is rendered in \autoref{fig:designspace}(c). \textit{directional layout} applies fade-in and fade-out to indicate the start and end of eye contacts in a smooth transition.

\autoref{fig:designspace}(d), (e), and (f) demonstrate how \textit{text overlay}, \textit{color highlights} and \textit{icon overlay} visualize eye contact indirectly. \autoref{fig:designspace}(d) shows text overlays at the bottom of video window. The names of observers are displayed following FIFO rule. The first name in the list is the one who looks at the observee earliest. \autoref{fig:designspace}(e) renders different color borders when different participants are looking at others. Likewise, \autoref{fig:designspace}(f) illustrates the profile at the bottom of the video window. The profile is a thumbnail image of the corresponding user. We take the first frame of video as a thumbnail reference. 

2D flat layouts are widely adopted in commercial video conferencing software. We provided two levels of eye contact visualization: direct and indirect. Direct eye contact options demonstrate eye contact to users intuitively, so it helps users immediately understand gaze information on a subconscious level. Indirect eye contact options imply the eye contact in a subtle way so users need to interpret the UI elements while the visual effects of elements are minimized so as not to be distracting.

% immersed?
\subsubsection{3D immersed layout}
We further investigate providing eye contact on 3D immersed layouts. The 3D immersed layouts are proposed to introduce spatial cues between participants as well as gaze information. The video image of all participants is attached to a monitor frame per user in the view. Instead of placing the video image directly in the scene and applying perspective transformation, we employ the ``physical kinetic displays''\cite{otsuka2016mmspace} metaphor and attach the video image to a monitor frame. This helps users to perceive the video as a 3D display. Other alternatives such as painting frames or mirror frames can also replace the monitor frames. We also designed the 3D immersed layout with two different perspectives: first-person view and third-person view. \autoref{fig:designspace}(g) and (i) are first-person perspective designs. In \textit{perspective layout}, users see a billboard representation of the video stream shaking if being looked at or turning if looking at other participants. Thus, we see user ``panda'' is shaking slightly in ``fox''s view and other users are turning to look at ``panda'' in \autoref{fig:designspace}(g). In \textit{avatar/first-person}, users see other participants as avatars representing their heads. The avatar will turn to the corresponding user when the user behind the webcam is looking at that user. So avatar `panda' is facing to the viewer (`fox') and other avatars are turning to look at `panda' in \autoref{fig:designspace}(i).

\autoref{fig:designspace}(h) and (j) show the third-person perspective designs in the 3D immersed layouts. In \textit{avatar / top-view}, all participants' avatars are rendered from above and a real-time video texture is shown alongside. Users can infer the spatial relationships from the orientation of the avatars. In \textit{avatar / third-person}, users' cameras are placed behind their own avatar so that each user can see other participants' head orientation as well as their own gaze cues.

The 3D immersed layout is designed to emphasize the spatial cues between participants. The effect is similar to 3D collaborative gaming experiences. We provide two levels of perspective and introduce 3D personalized avatars for user representation. Avatar representation is personalized according to real-time video textures (detailed in next section). The spatial cues and gaze information is designed to be natural and similar to real-life scenarios, though it may introduce some sense of the uncanny valley when warping the video image to fit the 3D avatars.

\subsubsection{hybrid layout}
We next explore the potential of hybrid layout designs that combine the 2D flat layout and 3D immersed layout. The hybrid layout is investigated to show a large video image and provide 3D gaze cues as well. We consider two rendering approaches: \textit{split-view} (\autoref{fig:designspace}(k) and \textit{picture in picture} (\autoref{fig:designspace}(l)). \textit{split-view} organizes the 2D layout and 3D layout side-by-side. Users are able to perceive eye contact from the 3D layout while simultaneously viewing the other participants in the 2D layout. \textit{picture in picture} allows users to acknowledge spatial cues at the center of their entire view. \textit{split-view} allows users to choose the focus on either video texture or eye contact and spatial information. \textit{picture in picture} prefers to present both pieces of information as a whole to users.

% how easy to see this information while watching the video / observing the other user.
% icon / text, how easy it is to relate the profile icon to the video stream.
% whether it is direct mapping.

% how easy it is to avoid uncanny valley
% how easy it is to scale to more people in the group

\subsection{Directional Layout and Perspective Layout}
% Why we choose these two to evaluate and show real screenshots
% how easy to see this information while watching the video / observing the other user.
% icon / text, how easy it is to relate the profile icon to the video stream.
% whether it is direct mapping.

% how easy it is to avoid uncanny valley
% how easy it is to scale to more people in the group
As the first step towards visualizing eye contacts for remote small-group conversations, we chose to implement and experiment with three conditions: baseline layout, 2D \textit{directional layout} and 3D \textit{perspective layout}. We have several considerations for selecting \textit{directional layout} and \textit{perspective layout} for comparison.
Our overarching goal is to explore how gaze and spatial information facilitate video conferences. \textit{directional layout} and \textit{perspective layout} both show eye contact in a direct manner, so that it is easy and straightforward for users to see and understand the system without further cognitive load. In addition, we chose not to include avatar designs in the first experiment because it would have required higher graphics processing capabilities of users' computers than designs not including avatars.
Also, it is likely that the personalized avatars could introduce ``uncanny valley'' effects, which might negatively impact users' conversation quality.
Lastly, hybrid layout designs are not selected for our exploratory user study, since we primarily want to understand how 2D flat layout and 3D immersed layout individually work for users.

\subsection{Use Cases}
% describe the scenarios and how each design adapted to different scenarios
\autoref{fig:designspace} demonstrates all the designs in a small-group discussion scenario. Meanwhile, video conferences are widely utilized in a large variety of use cases. LookAtChat is designed to be easily adapted to different video conferencing requirements.

\subsubsection{Small-group discussion}
\begin{figure}[t]
    \centering
    \includegraphics[width=\textwidth]{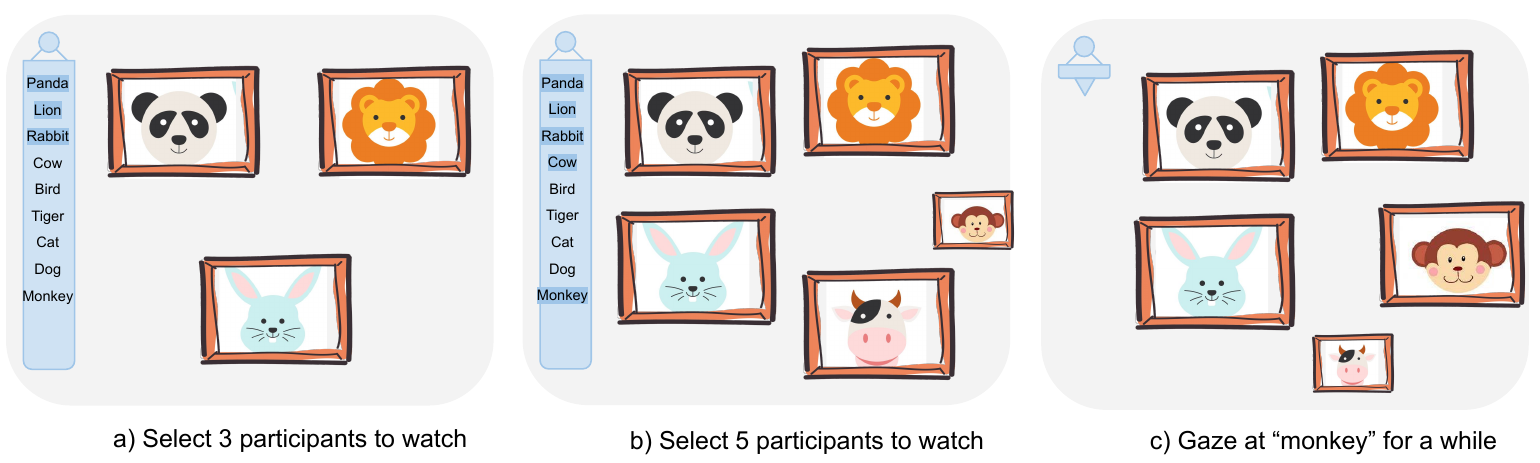}
    \hfill
     \caption{Video conferencing with varied numbers of participants in LookAtChat in 2D flat layout category. Size and placement of video image is updated in real-time according to the observer's gaze.}\label{fig:usecase:2D}
\end{figure}

% TODO: quantify a few participants of interest.
Small-group discussion is one of the majority use cases in video conferencing, either for working or for entertainment purposes such as brainstorming, playing games, etc. To save network bandwidth and decrease the cognitive load of users, LookAtChat provides a docked sidebar to show a full list of all participants. Users are free to choose a few participants of interest. The selected participants are rendered in a medium size video image at the beginning. The size will grow or shrink depending on the user's focus. Users can select or de-select participants at any time during the video conference (shown in \autoref{fig:usecase:2D}(a) and (b)). The participants who received focus from the user are gradually moved to the center \autoref{fig:usecase:2D}(b) (c).  

\subsubsection{Presentation}
Slides presentation is another strong use case in video conferences. During presentations, presenters focused mostly on the slides or the shared windows and on feedback from other participants. From formative interviews, people report being more interested in watching the slides than in watching the presenter.
So LookAtChat may visualize participants' focus with heatmaps.
Presenters will see other participants' gaze positions on the shared screen and listeners will see presenter's focus instead. 

\subsubsection{Large-scale meeting}
\begin{figure}[t]
    \centering
    \includegraphics[width=\textwidth]{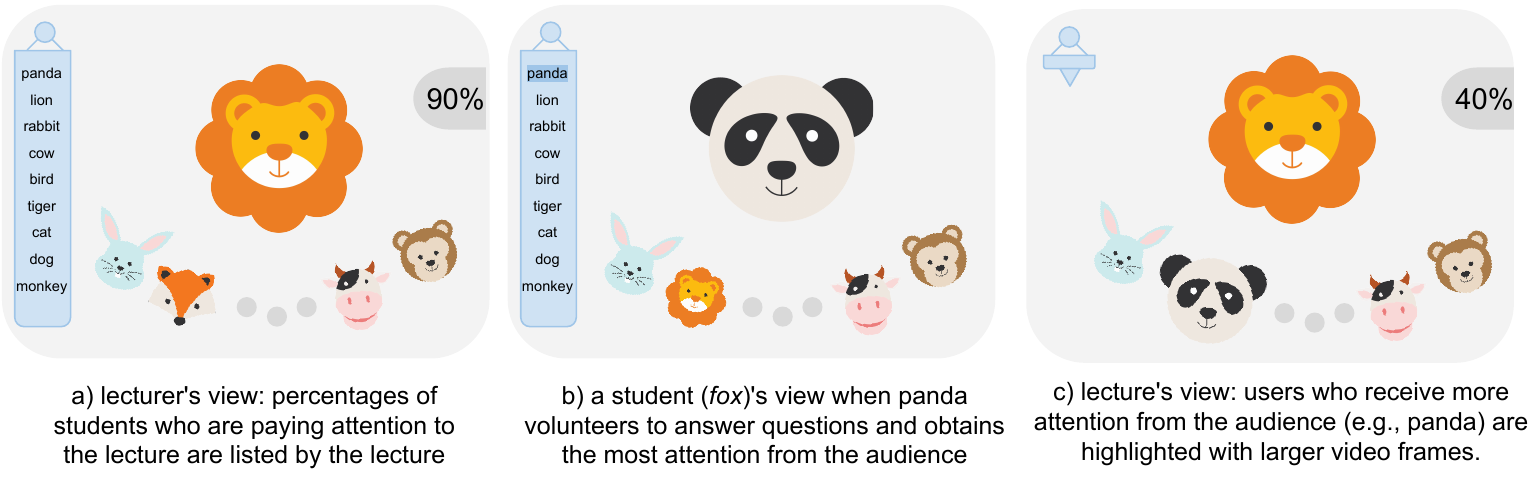}
    \hfill
     \caption{Presentation with a large audience in LookAtChat. An aggregated number of gaze-received is shown to the lecturer (see the top-right percentages in a) and c)). b) Audience members can choose to watch others in a large view. Audience members who receive more eye contact have larger size video frames than others for the lecturer to pay attention to.}\label{fig:usecase:3D}
     % TODO: make panda lion etc. non-capitalized if we have time; otherwise keep it the same.
     % -> a) lecturer's view: percentages of students who are paying attention to the lecture are listed by the lecture  b) a student (italic)'s view when panda volunteers to answer questions and obtains the most attention from the audience c) lecture's view: users who receive more attention from the audience (e.g., panda) are highlighted with larger video frames.
\end{figure}

Videoconferencing with a large audience is popular and useful for remote seminars and all-hands meetings. By default, the lecturer is rendered to all the participants (\autoref{fig:usecase:3D}(a)). Participants are free to see other participant as shown in \autoref{fig:usecase:3D}(b). We proposed to use aggregated data in this form. For example, \autoref{fig:usecase:3D}(a) shows how lecturers perceived other participants' looking at themselves and \autoref{fig:usecase:3D}(c) indicates when participants other than the lecturer receive focus. 

% use 2D icon without frame

% \subsection{Design Limitation}

\section{Implementation}

LookAtChat is designed and implemented for both video conferencing users and researchers to conduct remote user studies. LookAtChat comprises three major parts: a WebRTC server to support videoconferencing and hosting remote user studies, a real-time eye-contact detection module, and a WebGL-based renderer to visualize the gaze information.

\subsection{Workflow}
% host mode -- host user study

\begin{figure}[t]
    \centering
    \includegraphics[width=\textwidth]{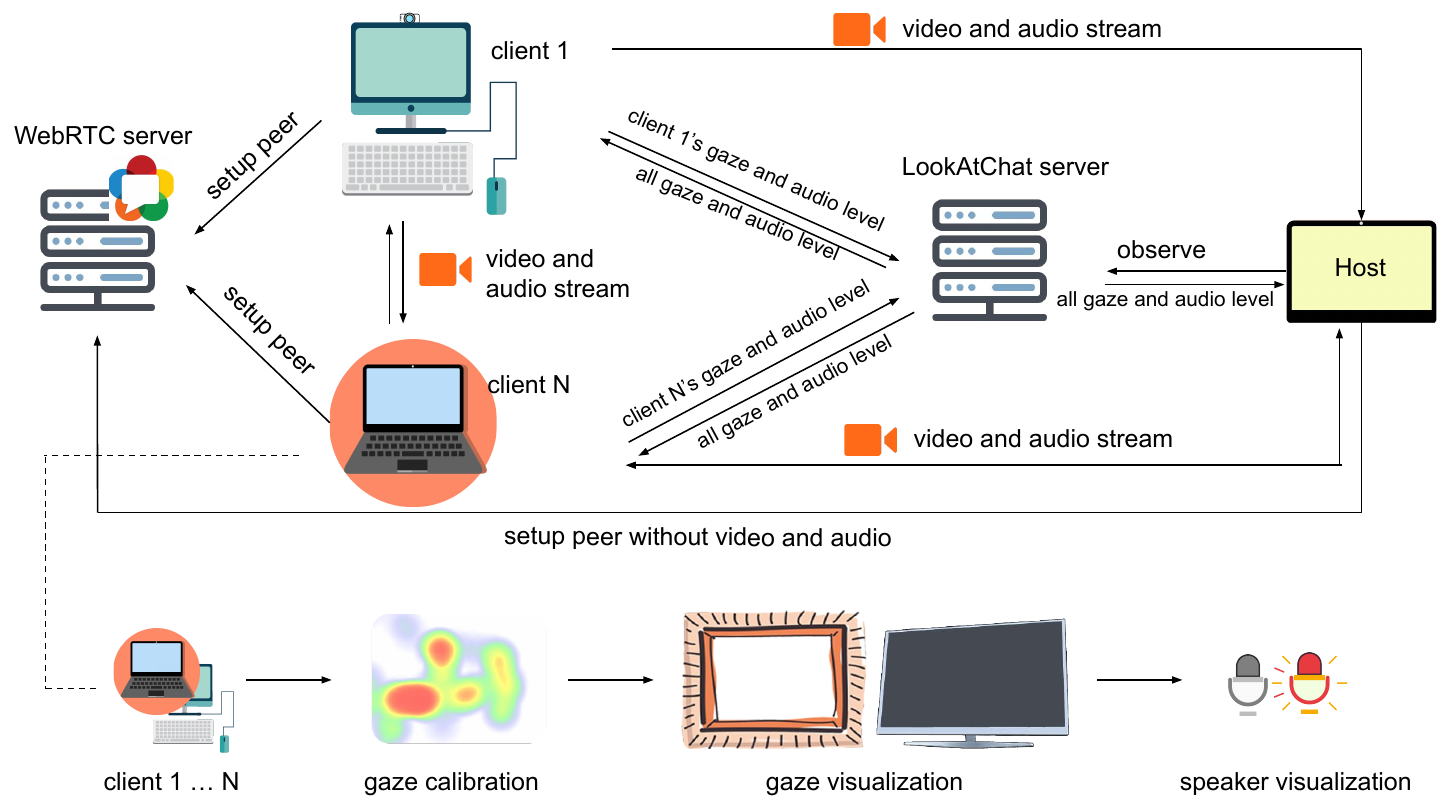}
    \caption{LookAtChat workflow. As a regular client, video and audio streams are transmitted to each other in peer-to-peer manner. Each client is required to calibrate gaze first and then send individual gaze and audio levels to the LookAtChat server after joining the group. The LookAtChat server broadcasts received data to every client. Then each client renders the gaze according to the current layout and the microphone symbol.} 
    % TODO: Client 1.. N -> clients 1 .. N; Calibrate gaze -> gaze calibration; Render gaze -> gaze visualization; Render microphone -> speaker visualization. Also we don't need to capitalize most words in the chart as they are NOT sentences.
    \label{fig:workflow}
\end{figure}

As \autoref{fig:workflow} demonstrates, LookAtChat employs a WebRTC server as well as peer-to-peer networking. For each newly-joined client, it talks to the WebRTC server (including Internet Connectivity Establishment server and Signaling) first to establish peer-to-peer connection with existing clients. Hence, the clients can send and receive video and audio streams with each other. Next, LookAtChat server maintains the identifier of each client after the WebRTC connection is established. For each client, gaze and audio level information are processed locally and sent to LookAtChat server. Afterwards, the server broadcasts the information to all of the clients and the renderer on the client side will locally visualize the gaze and audio information. 

\subsection{Host Mode to Support Remote User Study}

Due to the challenges of the global COVID-19 pandemic, it is not encouraged to recruit and gather participants in a controlled lab environment. Hence, we implement a \textbf{host mode} to monitor different clients from their own perspective and record gaze and video data. The host is a special client that does not participate in the actual study but can act as one of the participants for assisting them with technical issues. By default, the renderer will visualize all participants together with their eye contacts. The host can observe any client by sending an ``observe X'' command to the LookAtChat server. The server then returns the layout of all video streams as observed by the designated user. The renderer will visualize that user's gaze information so that the host can verify that the eye tracking modules are correctly calibrated.

To minimize bandwidth, the host does not send video and audio streams to other clients but only receives the streams from others. For post-study analysis, we record and save all the video, audio, and gaze data from the host machine.

Traditional video conferencing tools may allow the user to privately share their screens with the host. However, our approach reduces unnecessary communication efforts and allows the host to asynchronously examine remote user setups.

% describe the website directly
% TODO: need to add an index html to show three conditions

\subsection{Detection of Eye Contacts}
% webgazer
% accuracy with tobii
We leverage WebGazer~\cite{papoutsaki2016webgazer} to calibrate and obtain raw gaze positions in each client. Constrained by the webcam setup with ordinary laptop or PC, WebGazer is a state-of-the-art, off-the-shelf software for eye tracking technologies. Different from WebGazer, which reports gaze coordinates, LookAtChat focuses on ``who is looking at whom'' for video conferencing. We smooth the coordinate outputs from WebGazer with 1€ Filter~\cite{casiez20121} and then classify the data to understand which client is being looked at.

% TODO: face ratio
% I just measured, it is around 25%
Our system expects users to reach an accuracy of 80\% during the calibration session and ensure that the size of the face is larger than 25\% of the video frame. To improve accuracy, we fine-tune the parameters of 1€ Filter and video stream placement with a Tobii eye tracker.

%We ran a simulation to tune the above parameters. A Tobii tracker bar is connected as the ground truth. Horizontal parameters are independent from vertical ones. Thus, we applied the same procedure twice for horizontal and vertical parameters. 

% mincutoff = 0.3, beta = 0.3, dcutoff = 0.005
First, we tune parameter \textit{mincutoff} in 1€ filter to ensure the gaze coordinates are not jittering and \textit{beta} to ensure the result is not introducing too much latency. An animated dot moves from the start to the end of a line. The animated dot is a circle with a 10-pixel radius (from experimental data). We move our cursor to follow the dot several times and record the cluster of cursor positions. The sum of the average distance between cursor and animated dot is set as a reference value. We next move our gaze to follow the animated dot and apply the same calculation. \textit{Mincutoff} is tuned to ensure that the distance of gaze is comparable to the reference value. \textit{Beta} is tuned to ensure that the latency is shorter than 5 ms between the raw gaze position and smoothed gaze position on average. We set \textit{mincutoff} to be 0.3 and \textit{beta} to be 0.3 for smoothing the raw gaze positions. 
%Now we have smoothed WebGazer coordinates for further process.

% zone ID?
Second, we adjust the placement of the video stream. We calculate a zone ID to specify which video stream is being looked at. For ground truth (data from Tobii), we calculate a valid zone ID when the gaze dot is in the center of a video stream. The size of the central zone is defined as 1/4 of a regular video zone. With smoothed gaze coordinates, we calculate a valid zone ID when it is in the video zone. We reach 95\% after changing the distance between two video stream areas horizontally and vertically. Then the proportion of the distance between two video streams (both in $x$ and $y$) and the entire screen is recorded. In this way, we can ensure that LookAtChat behaves the same on different screens.

Eye contacts data is constructed as a pair of source ID and destination ID (which could be null) on each client. The data is sent to the LookAtChat server every 16 ms via web sockets to achieve real-time performance. % TODO: have we ever measure latency? if not, don't bother. As a reviewer, I would be curious about latency issues during the user study.

\subsection{Rendering}
\begin{figure}[t]
    \centering
    \includegraphics[width=\textwidth]{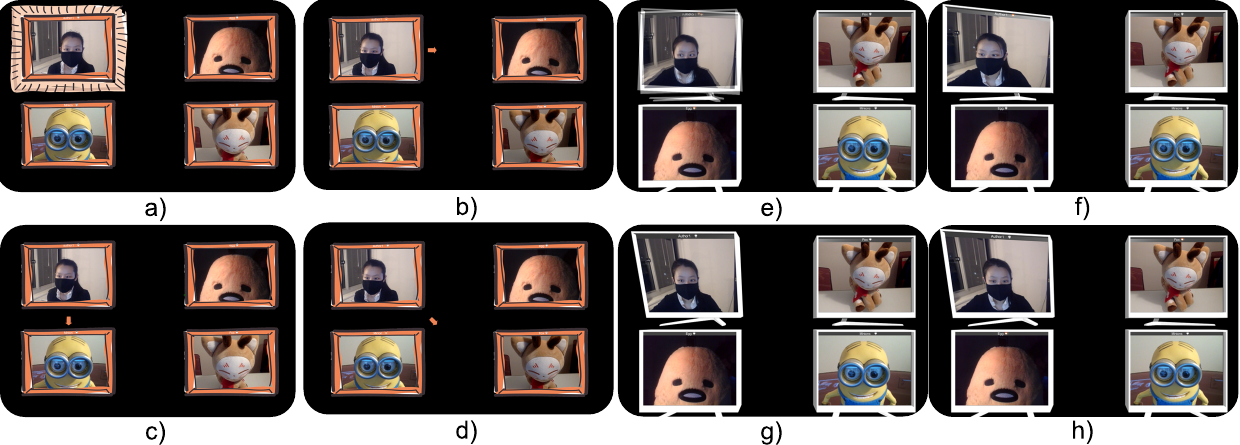}
    \caption{a) -- d) demonstrate screenshots of how the user recognize mutual gaze (a)) and eye contacts between the person on the upper left and other participants (b) -- d)) in the 2D directional layout. Accordingly, e) -- h) illustrate the gaze in the 3D perspective layout.} % TODO: I feel like g) is not fully looking downwards. 
    % emmm, see the upper and lower lines, they are parallel to each other.
    \label{fig:render:perspective}
\end{figure}

% TODO: LookAtChat renders two types of data on each client: A and B.
LookAtChat renders two types of data for each client. As we integrate WebRTC into our system for peer-to-peer video communication (including video stream and audio stream), the video stream is rendered as a video texture through Three.js. Additionally, we retrieve the local audio level on each client. The audio level is sent to LookAtChat server and broadcast to all the clients. Thus, each client can see a microphone icon ``on'' or ``off' based on the received audio level data (see ``author1'' in  \autoref{fig:render:perspective}(f)). We also record the audio level data for further data analysis.

Gaze data is rendered differently according to the layout. For directional layout, the out-glowing effect and the arrows are rendered with increasing opacity. Users can feel the action dynamically from the fade in and out effects. \autoref{fig:render:perspective}(a) to (d) show the view of the same user ``self'' in directional layout. For example, the video stream of user ``author1'' in (a) is outglowing because user ``author1'' is looking at user ``self''.
% TODO: slerp is just interpolation, we need to be clear how rotation/quaternion is applied
Regarding perspective transformation, the video image of the gaze source is transformed to facing to the video image of the target so that users can feel the movement of [who] is tuning and looking at [whom]. If the viewer is being gazed at, video texture of the gaze source will be slightly shaken. As \autoref{fig:render:perspective}(e) illustrates, user ``author 1'' is looking at ``self''. Accordingly, user ``author1'' is looking at ``fox'' on the right (\autoref{fig:render:perspective}(f)), at ``egg'' underneath (\autoref{fig:render:perspective}(g)), and ``minions'' at the corner (\autoref{fig:render:perspective}(h)). Interpolation is applied between different transformations for a smooth experience and also to simulate a ``turning'' action.

\section{Evaluation: Small-group Discussion}

We conducted a user study in order to examine how the different layout variants perform in terms of conversation, subjective feedback, and user preferences, compared with a ``baseline'' layout where no gaze information is visualized. The user study follows a within-subject design in three conditions: baseline, directional layout, and perspective layout. The three conditions were counterbalanced to avoid bias in the following combinations: baseline--directional--perspective, directional--perspective--baseline, perspective--baseline--directional. The DVs were conversation experience defined in Sellen's work ~\cite{sellen1995remote}, user experience defined in Schrepp \etal's work ~\cite{schrepp2017construction} and Hung \etal's work ~\cite{hung2017assessing}, and Temple Presence Inventory (TPI)~\cite{lombard2009measuring}. We processed the data through an analysis of variance (ANOVA). All tests for significance were made at the $\alpha=0.05$ level. The error bars in the graphs show the 95\% confidence intervals of the means.

% \subsection{Study Design}
% within subject
% data and analyze

\subsection{Participants and Apparatus}

We recruited a total of 20 participants at least $18$ years old with normal or corrected-to-normal vision (6 females and 14 males; age range: $24-39$, $M=28.55$, $SD=3.62$) via social media and email lists. The participants have a diverse background from both academia and industry.  None of the participants had been involved with this project before. We assign participants into four 5-person groups for the user study. 
The study was conducted remotely in personal homes.
Participants used their personal computers with a webcam, visited the website we provided through Google Chrome browser, and experienced different conditions as instructed by the host.
We instructed participants to take the user study in a quiet and brightly lit room where faces in the webcam are clearly visible from the background.
For the duration of the study, participants’ behavior, including their conversations, video streams, and gaze positions were observed and recorded.

\subsection{Procedure}

Our remote user study is scheduled using conventional calendar and videoconferencing tools (Zoom). Once all participants were online, the host briefly introduces LookAtChat system with a tutorial video and asks all participants to fill in consent forms. After the tutorial, the host instructs all participants to enter a designated layout in the user study website (\url{https://eye.3dvar.com}). Participants are instructed to mute their video\&audio streams in Zoom to prevent echoing and save networking bandwidth. Meanwhile, the participants can still follow the host's instructions from Zoom and the host can monitor the experiments with the \textbf{host mode} in LookAtChat. 
The user study session of each condition consists of three parts: gaze calibration ($\sim$5 min), warm-up conversation ($\sim$3 min), and two game sessions of ``who is the spy'' ($\sim$20 min). We now describe the three parts in more detail: 

\textbf{Gaze calibration}. 
Participants are required to first calibrate their gaze individually. Our system adopts the calibration procedure of WebGazer\cite{papoutsaki2016webgazer}: A box rendered around participant's face mesh turns green when the participant is at the center of the camera view and close enough. Next, the participant calibrates 9 points on the screen and the accuracy of gaze point is reported. We suggest that the participant proceed after reaching 80\%.

\textbf{Warm-up conversation}. In the beginning of each condition, researchers briefly describe how gaze information is visualized for the current condition. Later on, participants pick a topic (self introduction, favorite TV show, etc.) and one by one give a short speech for around 30 seconds. During the warm-up conversation, participants get familiar with the behavior of current condition.

% TODO: any other paper used who is the spy for games?
% nope. ppl discuss on a topic (MMSpace) to 2T1F
\textbf{Game ``who is the spy''}. After the warm-up conversation, the group is instructed to play a party word game, \textit{``Who is the spy''}, twice. Before the game starts, each participant receives a word: three of them act as detectives and get the same word ({\em e.g.}, William Shakespeare), while the other two act as spies and get a different word ({\em e.g.}, Leo Tolstoy). Only the spies know everyone's identities. For each round, each player needs to describe their word, talks about who may be the spies that received a different word. Spies will try to guess detectives' word and pretend they are holding the same word. Detectives will try to describe with ambiguity and infer spies with language and non-verbal cues. This game was selected because it is a conversation-based game which typically requires lots of eye contacts to tell who is lying and which two spies are teammates. At the end of each round, each player casts their vote for spy and the player with the most votes is put out of next rounds. At any time, detectives who successfully indict the spies and spies who successfully guess the detectives' word earn points. Each player has only one chance for the indictment or guess.

At the end of each study session, we ask the participants to fill an online questionnaire about conversation experience, user experience, and TPI~\cite{lombard2009measuring} with a 7-point Likert scale for the condition they just completed ($\sim$5 min). Hence, the study session of each condition lasts for around 30 to 45 minutes. At the end of all the three conditions, we ask the participants to fill demographic information, scale of usability in general, and rank the conditions. Lastly, participants are interviewed about LookAtChat, reasons for their ranking, and gave suggestions. On average, the experiment takes about 100 to 120 minutes in total. 

% purely about conversation and participants need to observe other participants. Spies It is an example that requires no other tools (white boards, screen sharing, etc) for small-group discussion.

\section{Results}
% we should report normality test result here
We validated that the data satisfies the assumptions of an analysis of variance (ANOVA). All tests for significance were made at the $\alpha=0.05$ level. The error bars in the graphs show the standard error. Symbol $^*$ means $p <= .05$, $^{**}$ means $p <= .01$, and $^{***}$ means $p <= .001$.

\subsection{Social experience}

\begin{figure}[t]
    \centering
    \includegraphics[width=\textwidth]{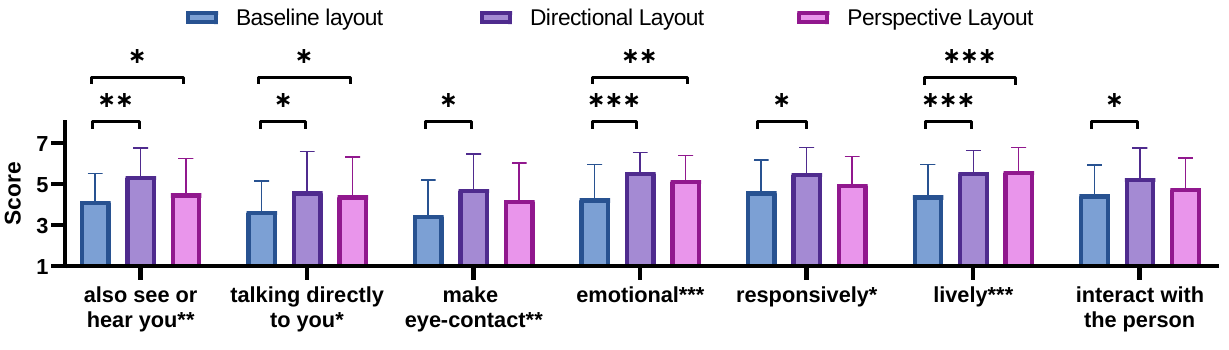}
    \caption{Summary of significant results regarding TPI between baseline layout, directional layout, and perspective layout. *: p <= 0.05, **: p <= 0.01, ***: p <= 0.001. 10 questions were selected in the category to be asked. We found 6 of them have significant effects through ANOVA and 7 of them have significant impacts in post hoc tests. Directional layout has notably better scores than baseline layout while perspective layout is slightly better.}
        % TODO: See how comprehensive our figure captions are in https://duruofei.com/papers/He_CollaboVR-AReconfigurableFrameworkForMulti-UserToCommunicateInVirtualReality_ISMAR2020.pdf
        % CollaboVR’s ratings, degree of helpfulness users in performing tasks, in synchronizing with partners, and in connecting with partners using the integrated layout (C1), mirrored layout (C2), and projective layout (C3). ∗ : p < 0.05, ∗∗: p < 0.01. We found a significant difference in ratings between C2 and C3; in degree of helpfulness between C1 and C2, C2 and C3; in synchronizing with partners between C1 and C2, C2 and C3. In terms of feeling connected with partners while using CollaboVR, the statistical results differed significantly among the three conditions. However, we did not find significant differences between each pair of conditions from post hoc tests.
    \label{fig:result:social}
\end{figure}

The results for the ratings of social experience questions that have significant results over three layouts are illustrated in \autoref{fig:result:social}.
% How often did you have the sensation that people you saw/heard could also see/hear you? 
For the question \textit{``How often did you have the sensation that people you saw/heard could also see/hear you?''} ($M_{baseline}=4.2, M_{dir}=5.4, M_{per}=4.55$), a one-way within-subjects ANOVA was conducted to test the influence of layout on the ratings. The main effects for layout ($F(1,19)=5.94, p=.006^{**}$) were significant. Post hoc t-tests with Holm correction showed a significant difference between baseline layout and perspective layout ($t(19)=-3.35, p<.006^{**}$) with a `large' effect size (Cohen's d $=.75$), as well as between directional layout and perspective layout ($t(19)=2.37, p<.046^{*}$) with a `medium' effect size (Cohen's d $=.53$). The results indicate that LookAtChat with directional layout and perspective layout provided notably more bidirectional sensation than baseline layout.

For the question \textit{``How often did it feel as if someone you saw/heard was talking directly to you?''}($M_{baseline}=3.7, M_{dir}=4.65, M_{per}=4.45$), a one-way within-subjects ANOVA was conducted. The main effects for layout ($F(1,19)=4.93, p=.012^{*}$) were significant. Post hoc t-tests with Holm correction showed significant differences between baseline layout and other two layout (both $p<.05^{*}$) with a `medium' effect size (Cohen's d $=.52$ to $.66$). The results suggest that LookAtChat with directional layout and perspective layout provided notably more feelings of direct conversation than baseline layout.

For the question \textit{``How often did you want to or did you make eye-contact with someone you saw/heard?''}($M_{baseline}=3.5, M_{dir}=4.75, M_{per}=4.25$), a one-way within-subjects ANOVA was conducted. The main effects for layout ($F(1,19)=6.71, p=.003^{**}$) were significant. Post hoc t-tests with Bonferroni correction showed significant differences between baseline layout and directional layout ($p<.002^{**}$) with a `large' effect size (Cohen's d $>.8$). The results indicate that LookAtChat with directional layout provides significantly more eye contact than baseline layout.

Regarding the social richness questions including emotional ($M_{baseline}=4.3, M_{dir}=5.6, M_{per}=5.2$), responsive ($M_{baseline}=4.7, M_{dir}=5.6, M_{per}=5.0$), and lively ($M_{baseline}=4.5, M_{dir}=5.6, M_{per}=5.7$), a one-way within-subjects ANOVA was conducted for each of them. 
The main effects for layout were significant on ``emotional'' (($F(1,19)=11.13, p<.001^{**}$)) with post hoc t-tests with Bonferroni correction showed significant differences between baseline layout and the other two layout (both $p<.006^{**}$) with a `large' effect size (Cohen's d $>.71$); 
on ``responsive'' (($F(1,19)=3.69, p=.03^{*}$)) with post hoc t-tests (holm correction) showed significant difference between baseline layout and directional layout (both $p=.03^{*}$) with a `medium' effect size (Cohen's d $=.6$); 
on ``lively'' (($F(1,19)=10.65, p<.001^{**}$)) and post hoc t-tests with Bonferroni correction showed significant differences between baseline layout and the other two layout (both $p<.001^{***}$) with a `large' effect size (both Cohen's d $>.87$). The results indicate that LookAtChat with directional layout and perspective layout provides significantly more social richness feelings than baseline layout.

For other questions discussing social experience, we did not find significant effects over the three layouts. Furthermore, we found that question \textit{``to what extent did you feel you could interact with the person or people you saw/heard?''} has ``large'' effect size ($\eta^2=.144$). Post hoc t-tests with Holm correction shows baseline layout has significantly negative effects ($p<.05^{*}$) compared with the directional layout. Hence, the result suggests that LookAtChat with directional layout brings more interaction potential to users than the baseline layout.

\subsection{User engagement and experience}
\begin{figure}[t]
    \centering
    \includegraphics[width=\textwidth]{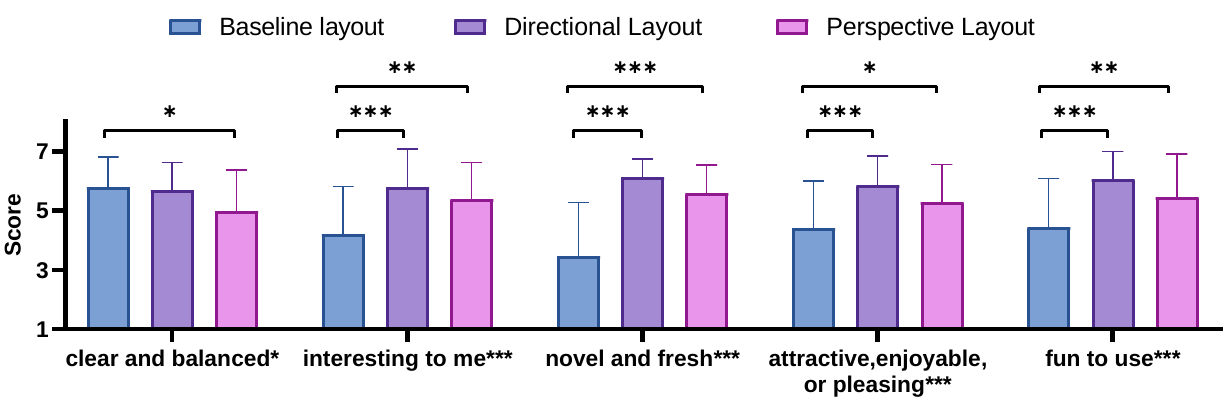}
    \caption{Summary of significant results regarding user engagement and user experience between baseline layout, directional layout, and perspective layout. *: p <= 0.05, **: p <= 0.01, ***: p <= 0.001. We provided 6 questions on user engagement and 4 questions on user experience for participants to fill. Results show that 3 user engagement questions and 2 user experience questions have significant effects through ANOVA. Directional layout and perspective layout both have significant effects on participants' feedback of feeling ``interesting'', ``novel'', ``attractive'' and ``fun''.}
    \label{fig:result:ux}
\end{figure}

The ratings of the question \textit{``The visualization of the layout is clear and balanced.''} over three layouts are illustrated in Figure~\ref{fig:result:ux} ($M_{baseline}=5.8, M_{dir}=5.7, M_{per}=5.0$). A one-way within-subjects ANOVA was conducted to test the influence of layout on the ratings.
The main effects for layout ($F(1,19)=3.83, p=.03^{*}$) were significant. Post hoc t-tests with Holm correction showed a significant difference between baseline layout and perspective layout ($t(19)=2.54, p<.046^*$) with a `medium' effect size (Cohen's d $=.57$). The results indicate that LookAtChat with baseline and directional layout is more clear and balanced than perspective layout.

Regarding \textit{``the content or features provided on this website were interesting to me.''} ($M_{baseline}=4.2, M_{dir}=5.8, M_{per}=5.4$), the main effects for layout ($F(1,19)=9.15, p=.001^{***}$) was significant with a `large' effect size ($\omega^2=0.18$). Post hoc t-tests with Holm correction showed a significant difference between baseline layout and directional layout ($t(19)=-4.11, p<.001^{***}$) and perspective layout ($t(19)=-3.08, p<.008^{**}$). The results indicate that LookAtChat with baseline layout is significantly less interesting than directional and perspective layout.

In terms of \textit{``the features provided by this website were novel and fresh.'' } ($M_{baseline}=3.45, M_{dir}=6.15, M_{per}=5.6$), the main effects for layout ($F(1,19)=32.76, p<.001^{***}$) were significant with a `large' effect size ($\omega^2=0.47$). Post hoc t-tests with Holm correction showed a significant difference between baseline layout and directional layout and perspective layout (both$p<.001^{***}$). The results indicate that LookAtChat with baseline layout is significantly less novel or fresh than directional and perspective layout.

Speaking of the overall impression of the design (attractive, enjoyable, or pleasing) with $M_{baseline}=4.4, M_{dir}=5.9, M_{per}=5.3$. The main effects for layout ($F(1,19)=9.28, p=.001^{***}$) were significant with a `large' effect size ($\omega^2=0.16$). Post hoc t-tests with Holm correction showed a significant difference between baseline layout and directional layout ($t(19)=-4.3, p<.001^{***}$) and perspective layout ($t(19)=-2.6, p=.02^{*}$).
The result for ratings of \textit{``fun to use''} over three layouts are $M_{baseline}=4.5, M_{dir}=6.1, M_{per}=5.5$. The main effects for layout ($F(1,19)=11.49, p=.001^{***}$) were significant with a `large' effect size ($\omega^2=0.17$). Post hoc t-tests with Holm correction showed a significant difference between baseline layout and directional layout ($t(19)=-4.75, p<.001^{***}$) and perspective layout ($t(19)=-2.97, p=.01^{**}$). The results suggest that LookAtChat with the baseline layout is significantly less attractive and fun than the directional and the perspective layouts overall.

\subsection{Conversation experience}
\begin{figure}[t]
    \centering
    \includegraphics[width=\textwidth]{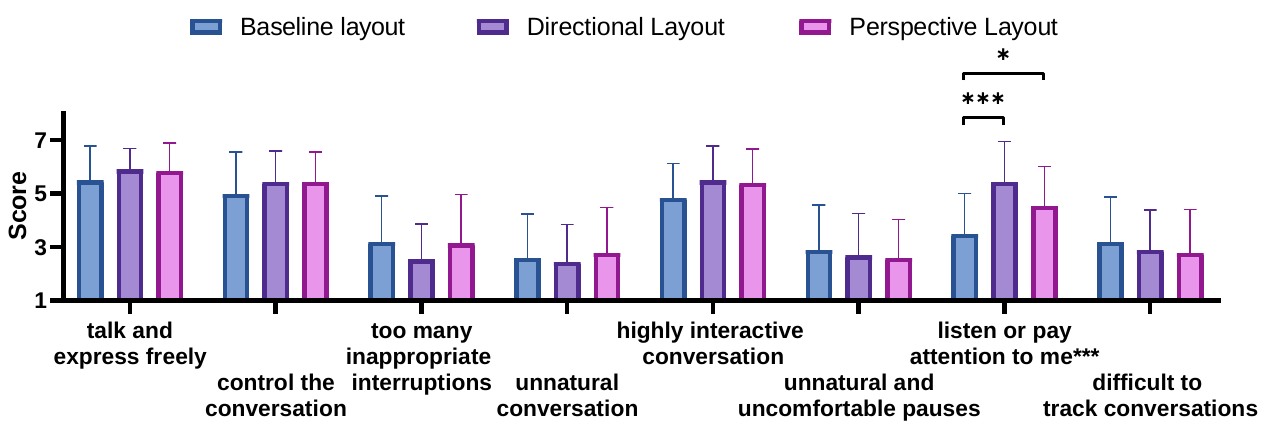}
    \caption{Summary of the results regarding conversation experience between baseline layout, directional layout, and perspective layout. *: p <= 0.05, **: p <= 0.01, ***: p <= 0.001. Directional layout and perspective layout both have significant effects on the feeling of ``attention''. For other questions, directional layout and perspective layout have better but not significant scores than baseline layout.}
    % TODO: See how comprehensive our figure captions are in https://duruofei.com/papers/He_CollaboVR-AReconfigurableFrameworkForMulti-UserToCommunicateInVirtualReality_ISMAR2020.pdf
    \label{fig:result:conv}
\end{figure}

% I knew when people were listening or paying attention to me.
The results for the ratings of question\textit{ ``I knew when people were listening or paying attention to me.''} over three layouts are illustrated in \autoref{fig:result:conv} ($M_{baseline}=3.5, M_{dir}=5.45, M_{per}=4.55$). A one-way within-subjects ANOVA was conducted to test the influence of layout on the ratings.
The main effects for layout ($F(1,19)=12.50, p=.001^{***}$) were significant. Post hoc t-tests with Holm correction showed a significant difference between baseline layout and directional layout ($t(19)=-5.00, p<.001^{***}$) and perspective layout ($t(19)=-2.68, p<.02^*$) with a `large' effect size (both Cohen's d $>.6$). The results indicate that LookAtChat with both directional layout and perspective layout provided notably more feedback of attention than baseline layout.

We found there was no significant effect of layout on other conversation questions. Post hoc t-tests with Holm correction did not show significant differences among all questions in this category.

% \subsection{Structure of the conversation}
% need to go over the videos QAQ
% TODO

\subsection{Usability, Preferences and Subjective feedback}
Participants rated the usability of LookAtChat in general after they experienced all three layouts. Participants (like P10,M) thought the system is ``\textit{easy to understand}''. The overall usability score is 72.5, which is interpreted to have higher perceived usability than 70\% of all products tested.

Fourteen participants preferred the directional layout, five participants preferred the perspective layout, and one preferred the baseline layout. P6(M) thought ``\textit{arrows and highlights are easier for me to notice the other's attention}''. Similarly, P11(F) agreed that ``\textit{the highlight of when people look at you is more obvious.}''. For participants who preferred perspective layout, participants found it ``\textit{fun}''(P9, M) and ``\textit{having such eye contact design in video conferencing is interesting and I feel the motive of being focus on other's speeches.}'' (P7, F). Additionally, P1(M) and P2(F) shared similar comments that ``\textit{more intuitive than the other designs}'' and ``\textit{simple form but it also gives you info}''. The only user (P4,F) who preferred the baseline layout explained that the baseline layout is ``\textit{succinct}''.
From the reason of preferences reported by the participants, we found that participants tended to choose the design that is ``clear and easy to understand'' (P18, F) and (P12, F), however, they have different perceptions on defining ``clear''. Providing information that is more than users expect lead to negative results. An automatic adjuster or manual one would be useful.

During the interview, we asked all the participants whether they think eye contacts are worth visualizing as well as spatial relationship. 15 out of 20 agreed it is helpful and others have concerns about being overwhelmed. We further asked how they think about eye contacts between themselves and other people, and eye contacts between other people. 19 out of 20 reported they want to know if they are being looked at as well as whether other people receive their gaze information. We believe that visualizing eye contacts data will improved users' engagement, social richness, and helpful to conversation experience while videoconferencing. Since LookAtChat only requires a consumable webcam, participants such as P4(F) and P5(M) reported that it should be feasible to be integrated into existing videoconferencing systems, although may cause extra rendering costs.
% TODO others from interview part

\section{Discussion}
In general, LookAtChat is an \textit{``easy-to-use''} system with an overall usability score of 72.5 in this user study. Participants found it \textit{``easy to understand''} and \textit{``fun to use''}. Compared with the baseline layout, the directional layout and the perspective layout provide better feedback on social experience, user engagement, and conversation experience.

\textbf{RQ1: How do people perceive eye contact without visualization in conventional videoconferencing?}

Informed by the interview feedback, participants usually don't interpret eye contacts in videoconferencing. Commented by P18(F), ``\textit{I focus more on the audio so I won't miss what other people are talking about.}''. In the meantime, participants are used to interpreting gaze offset to a very high degree of accuracy~\cite{grayson2003you}. Hence, providing eye contacts through user interface is worth researching in parallel with gaze correction technologies.

\textbf{RQ2: Can visualization of eye contact improve communication efficiency?}

According to the conversation experience feedback, participants have better experiences with conversation flow, including self expression, controlling the conversation, and tracking the conversation, but not significantly. Participants felt the conversation is more interactive. Meanwhile, participants reported fewer inappropriate interruptions, fewer uncomfortable pauses, in addition to feeling less unnatural when using LookAtChat with directional or perspective layouts though not significantly. Importantly, LookAtChat has a significant effect on participants' belief that other people are listening and paying attention to them. As P4(F) reported, ``\textit{visualization of eye contact is definitely helpful, knowing that there were people watching at least helped me stay focused for the entire time.}''

\textbf{RQ3: In what circumstances will users prefer to see eye contact? }

We found that participants have different preferences while in different roles, as a meeting host, or as a talk attender. In addition, the purpose of the video conference mattered. As P5(M) described, ``business meetings may value more on the quality of video image however colorful UI designs (like arrows in directional layout) may distract people from that''. For small-group discussion, P4(F) agreed that ``providing such information motivates me to engage more in such videoconferencing like brainstorming.'' Furthermore, participants placed more value on eye contacts relevant to themselves than between other participants. P5(M) thought eye contacts between other participants ``\textit{are helpful but too much for me to process at one time}''. Investigating the effects of enabling eye contacts among other participants is worth researching. 

\textbf{RQ4: What forms of eye contact visualization may be preferred by users? }

More participants preferred the directional layout (N=14) than the perspective layout (N=5). 
The directional layout shows arrows and outer glows that ``\textit{are easier to notice other's attention}''(P6, M). It applies ``\textit{no change on video image}''(P18, F) and indicates the interaction ``\textit{more visually for better connection}'' (P13, M).
The perspective layout provides spatial relationships in 3D. Reported by P1(M), it has ``\textit{better visual presentation of gazing at somebody, more intuitive}''. P7(F) Comparing directional and perspective layout, ``\textit{directional layout is somehow too obvious and may require extra effort to focus on my speech}'' while ``\textit{perspective layout eliminates this issue and I got the balance between freely speaking and knowing I was listened to by others.}''. Briefly, we can tell that participants prefer the the designs that are ``clear'' to them and with smooth transition between looking at self and others.

\textbf{Design Implications:}
\begin{enumerate}
    \item Show the eye contacts intuitively. 
    \\In video conferences, participants mainly focus on the conversation. If the information provided through the visual design is too indirect and may require additional cognitive load, participants may feel distracted and lost in the conversation. For example, out-glowing in the directional layout receives positive comments from 7 participants because it is easy to understand.
    \item Control the visualization level. 
        % TODO: What does the first sentence mean? What does the level mean?
    \\Participants' preference is affected by ``how attractive the design is'' and ``how I want the design to be attractive''. For example, P7(F) preferred the perspective layout because the directional layout is relatively more distracting to her when she wanted to focus on listening. Providing a slider for adjusting the level and automatically controlling the level with audio data is helpful.
    \item Provide the control of gaze data transmission.
    \\ As video conferencing users have the options to mute or hide the video during video conferencing, most participants want to hide or send anonymous gaze data to the host in video conferences.
    \item Scale the design for various user scenarios.
    \\ Although we only evaluate small-group setup for LookAtChat, we designed and interviewed participants about their opinions on large team meetings or presentation with a large audience. An important future direction is to adapt the system to fit different use cases and larger numbers of users.
    \item Provide host mode.
    \\ Host mode is not only helpful for researchers to understand participants from their views but essential for conducting remote user studies as well.
\end{enumerate}

\textbf{Limitations}.
% TODO: I added more limitations to prevent reviewers to claim that we are not honest... Try to persuade them to look at this more positively.
While LookAtChat is designed for remote video conferencing, as a proof-of-concept, we do not support more than one participant to be co-located. Our eye-contact detection algorithm only supports one user in front of the webcam and the accuracy is limited by the algorithms and individual calibration procedure.
In terms of the user study, we only evaluate small-group discussion without shared presentation. As
the ages of our participants spanned 24 - 39, the results of our study may not generalize to other populations such as junior students or elder adults who may prefer more or less eye contacts in video conferences. As our user study was conducted remotely, the bandwidth of home networks may impact how the users perceive our system. A small number of participants encountered frame-dropping during the conversation due to networking issues, which may negatively impact their assessment. Furthermore, as users of our system were only able to engage with LookAtChat for casual and gaming conversations, their assessment of how such system may help in other use cases such as team meetings or lectures is not fully conclusive.

% Potential benefits and challenges on incorporating eye contact visualization into existing videoconferencing systems are also discussed. 

\section{Conclusion and Future Work}
In this paper, we introduce LookAtChat, a web-based video conferencing system which supports visualizing eye contacts for small-group conversations. Motivated by missing gaze information in conventional video conferences, we investigate the demands of gaze information by conducting five formative interviews. We further explore the design space of visualizing eye contacts with video streams of small groups and propose 11 layouts by brainstorming in focused groups. As a proof-of-concept, we develop and open source LookAtChat which supports eye contact visualization for small-group conversations. We conduct a remote user study of 20 participants to examine the benefits and limitations of the interfaces, as well as the potential impacts of user engagement and experience on the conversation. The quantitative results indicate that LookAtChat with directional layout and perspective layout provided notably more bidirectional sensation, feelings of direct conversation, social experience, and engagement than the baseline layout. More participants prefer the 2D directional layout to the 3D perspective layout because it is simpler and easier to understand.

We plan to explore several future directions for improving LookAtChat. First, we plan to implement more layouts from our design space exploration stage and establish an open-source community to develop more layouts to the system. Second, we intend to integrate privacy protection filters for users to select whether or not to share their own gaze information. Third, more advanced real-time neural models may be leveraged to improve the tracking accuracy in LookAtChat and balance the trade off between accuracy and real-time performance.

As an initial step toward visualizing eye contacts in conventional video conferencing interfaces with commonly accessible hardware requirements, we believe our work may inspire more designs to convey nonverbal cues for remote conversations. Such features may eventually be integrated with video conferencing software to increase social engagement and improve conversation experience.

% Potential benefits and challenges on incorporating eye contact visualization into existing videoconferencing systems are also discussed. 
% we believe that our system can greatly enhance social engagement for remote collaboration, laying down the groundwork for future systems to visualize nonverbal cues and accommodate richer interactions among larger groups. 

% \begin{acks}
% ...acknowledgments...
% \end{acks}

%%
%% The next two lines define the bibliography style to be used, and
%% the bibliography file.
\bibliographystyle{ACM-Reference-Format}
\bibliography{ACM-Reference-Format}

%%
%% If your work has an appendix, this is the place to put it.
% \appendix

% \section{...appendix...}

\end{document}